\documentclass[journal]{IEEEtran}

\usepackage{graphicx}
\usepackage{epsf}
\usepackage{amsmath, amsfonts}
\usepackage{latexsym}
\usepackage{multirow}
\usepackage{multicol}
\usepackage[table]{xcolor}

\PassOptionsToPackage{hyphens}{url}
\usepackage[unicode=true, bookmarks=true, bookmarksnumbered=true, bookmarksopen=true, bookmarksopenlevel=1, breaklinks=false, pdfborder={0 0 0}, pdfborderstyle={}, backref=false, colorlinks=false]{hyperref}

\usepackage{bbm}
\usepackage{subcaption}

\definecolor{green}{RGB}{0,170,0}

\begin{document}

\title{\fontsize{23}{2}\selectfont Human Electromagnetic Field Exposure\\\vspace{0.12 in}in Wearable Communications: A Review}
\author{\\\{seungmokim, ys01063\}@georgiasouthern.edu\\Department of Electrical and Computer Engineering\\Georgia Southern University}

\author
{
Seungmo Kim, \textit{Member, IEEE}, Yakub Ahmed Sharif, and Imtiaz Nasim

\thanks{S. Kim and Y.A. Sharif are with Department of Electrical and Computer Engineering, Georgia Southern University in Statesboro, GA, USA (e-mail: \{seungmokim, ys01063\}@georgiasouthern.edu). I. Nasim is with Department of Electrical and Computer Engineering, Florida International University in Miami, FL, USA.

The work shown in Section \ref{sec_mmw} of this paper was presented in this paper was presented in IEEE MILCOM 2019 \cite{milcom19}.}
}

\maketitle

\begin{abstract}
The concern on human health is often overseen while wearable technologies attract exploding interests. Mainly due to the extreme proximity or a direct physical contact to the human skin, wearable communications devices are acknowledged to cause higher levels of specific absorption rate (SAR) at the skin surface. Unfortunately, so far, we have found no study encompassing all the aspects that the general public needs to understand about wearable technologies--\textit{i.e.}, the analytical and experimental backgrounds, and report of SAR levels generated from commercial wearable devices. In this context, this paper provides an extensive review on SAR from various commercial wearable devices that are currently sold in the market, as well as the analytical framework and the current measurement methodologies for standard compliance tests. Moreover, considering the present interest in millimeter wave (mmW), this paper sheds light on the SAR evaluated at 60 GHz and also compares the SAR to that measured at 2.4 GHz. We expect that this paper will be of value in informing the general public of the safety in using the currently sold wearable devices, and in igniting further study of the exact biological consequences from electromagnetic field (EMF) exposure due to wearable devices.
\end{abstract}


\section{Introduction}
\subsection{Significance of Wearable Technologies}
Mobile wearable communications devices are rapidly making inroads thanks to advancements in miniature electronics fabrication, wireless communications, batteries, and data analytics. The initial driver of the mobile electronics market was fitness and healthcare gadgets \cite{wncg}. The convergence of wearables, such as smartwatches and activity trackers, has initiated the growth of smart technology in healthcare--\textit{e.g.}, wireless blood pressure/respiratory rate wristbands or patches for home diagnosis \cite{dias18}.

Today, we evidence a remarkable expansion of wearable technologies in the mobile device market \cite{gadget18}. September 2009 can be considered as the defining moment when iconic wearable technologies like Nike+ and Fitbit became commercialized \cite{easychair}, which further expanded the industry and led to an unprecedented variety of applications: \textit{e.g.}, smartwatches, wristbands, smart sunglasses, smart jewelry, electronic clothing, skin patches, etc \cite{carames18}. Data obtained from these wearables creates opportunities that will improve the quality of life that existing mobile technologies (\textit{e.g.}, smartphone) alone could not easily achieve \cite{seneviratne19}.

Not only for personal life, wearable devices are also used to provide a wider range of value-added services such as indoor positioning and navigation \cite{z. yang13}\cite{H. Lee149}, financial payments \cite{Apple Pay}\cite{Payment Technology}, physical and mental health monitoring \cite{Vidal35}\cite{Wijsman}, sports analytics \cite{Anzaldo}, and medical insurance analytics \cite{forbes}. Wearable devices can provide easier access to information and more convenience for their users. They have varying form factors, from low-end health and fitness trackers to high-end virtual reality (VR) devices, augmented reality (AR) helmets, and smartwatches \cite{vtmag18}.

\subsection{Health Concern on EMF Emitted by Wearable Devices}
A major concern regarding wearable communications is human biological safety under exposure to EMF generated by the wearable devices \cite{vtmag18}. A human body absorbs electromagnetic radiation, which causes thermal or non-thermal heat in the affected tissues. Further, as shall be informed in Section \ref{sec_analysis} of this paper, the EMF energy arriving at the human skin is dominated by the distance between the EMF emitting device and the skin. This makes wearable communications as the type of wireless technology on which the most careful analysis is needed regarding human EMF exposure.

For this reason, there has been surveys and tutorials on the human EMF exposure in wearable communications \cite{easychair}\cite{traj16}\cite{elsevier18}, which provided up-to-date information that is necessary to keep general consumers informed of safe use of wearable technologies. Despite the significant contributions, the prior surveys limit their generality due to lacking enough details on:
\begin{itemize}
\item Exact levels of EMF exposure caused by the commercial wearable products that are currently in the market
\item Potential impacts of technological change in future (\textit{i.e.}, use of mmW for wearables) on the EMF exposure
\end{itemize}

\subsection{Contribution of This Paper}
To this end, this paper provides an extensive overview on the human EMF exposure in wearable communications, addressing the aforementioned limitations of the prior surveys \cite{easychair}\cite{traj16}\cite{elsevier18}. The key contributions of this paper are highlighted as follows:
\begin{itemize}
\item \textit{Investigation on SAR Levels of Off-the-shelf Wearable Products:} The exact SAR levels generated by the popular commercial wearable devices--\textit{i.e.}, Apple AirPod, Fitbit, Snap wearable video camera, and Apple Watch--are investigated.
\begin{itemize}
\item The values have been found via (i) investigation of a product's data sheet and external articles, and (ii) calculation referring to the standard analysis and measurement frameworks in case a SAR level is not revealed.
\end{itemize}
\item \textit{State-of-the-art Standards and Measurement Methodologies:} This paper features in-depth information on how the SAR values shown in the products are computed.
\item \textit{SAR Analysis According to Spectrum Band:} An explicit comparison of SAR at different operating frequencies is provided--\textit{i.e.}, Wi-Fi, Universal Mobile Telecommunications System (UMTS), Long-Term Evolution (LTE), etc.
\begin{itemize}
\item Some products support multiple spectrum bands--\textit{e.g.}, Apple Watch series, as shall be discussed in Section \ref{sec_devices}. The contribution of this paper is highlighted in such a case: we lead the customers to precise understanding on differentiated impacts according to carrier frequencies.
\end{itemize}
\item \textit{SAR Evaluation of Wearables Operating in mmW Spectrum:} Section \ref{sec_mmw} compares the SAR caused by a wearable device in 2.4 GHz and 60 GHz.
\begin{itemize}
\item Wearable communications at 60 GHz adopt a more directional radiation pattern to (i) overcome the excessive path loss and (ii) achieve higher data rates at the user end compared to the previous-generation technologies \cite{milcom19}. Such directivity elevates the SAR on human skin.
\end{itemize}
\end{itemize}

\section{State-of-the-Art}\label{sec_stateoftheart}
This paper starts the survey with review on the current understanding in the literature. This section can suggest a taxonomy of the current study on human EMF exposure.

\subsection{Wearable Communications Technologies}
\subsubsection{Supporting Sensors}
An overview on available electronic components and toolkits to construct smart garments has been introduced \cite{Hartman}. The actual operating mechanisms of various sensors available for wearable devices and different fundamental algorithms associated with them (\textit{i.e.}, step counting and activity recognition) were also presented \cite{Sazonov}.

\subsubsection{Medical Applications}
Thus far, medicine has been the area to which the most various applications of wearable sensors have been implemented: \textit{e.g.}, health monitoring/prognosis \cite{A. Pantelopoulos40}-\cite{M. M. Baig51}, activity recognition \cite{O. D. Lara15}, and assisted living \cite{Dakopoulos40}.

Several interesting technical challenges have been studied. Wearable devices have limited visual and auditory data output channels due to their physical characteristics and complex use scenarios \cite{Wang18}. A comparative study of notification channels (light, vibration, sound, poke, thermal, etc.) for wearable interactive rings had an interesting conclusion regarding the challenge: vibration was found to be the most reliable and fastest in conveying sensor notifications \cite{Roumen}.

A wearable networking mechanism connected to a cloud has also been studied. A cloud-based mobile calorie monitoring system was proposed, where the system can categorize edible items in a dish and then process the complete calorie count with high accuracy, for each food item \cite{Peddi66}.

\subsection{Concern on EMF Exposure}
While mostly focusing on techniques and applications only, the literature of wearable communications is scarce in regards to human EMF exposure. Mainly due to the nature of its operation on physical contact on a human body, the wearable communications pose particular attention in the literature. As such, recent research has shown interest in the impact of SAR on the human body.

More than 240 scientists who have published research on electromagnetic radiation safety believe that current national and international guidelines for exposure to radio frequency radiation are inadequate to protect human health \cite{EMFscientist}.

For instance, a latest study compared the EMF exposure amounts in different spectrum bands \cite{milcom19}. The 2.4 GHz band was studied as a key spectrum band in which current wearable devices operate. Also, the study compares the SAR to that measured at 60 GHz, acknowledging that ``higher-rate'' Wi-Fi standards (\textit{i.e.}, IEEE 802.11ad and ay) operating in this band are enablers of the next-version wearable communications \cite{vtmag18}. It was found that the SAR at 60 GHz can exceed the regulatory guidelines within a certain separation distance between a wearable device and the human skin surface--\textit{i.e.}, 12 to 15 mm \cite{milcom19}.

\subsection{EMF Exposure in Commercial Wearable Products}
Hitherto, there has been no comprehensive survey investigating the EMF exposure in currently sold commercial wearable devices. According to a recent article \cite{readwrite}, many of fitness wear devices that are currently out in the market have been found to be untested, unscientific, and anxiety causing. Of course there are other opinions stating that the concerns are exaggerated and the EMF from earbuds is far too weak to have any noticeable impact on the human body \cite{bi19}. However, we believe that such a controversy only fortifies justification of further investigation and conservative deployment of high-energy-emitting wearable devices.

\subsubsection{Fitbit}
In fact, Fitbit, a company well-known for its Wi-Fi-based watch-type health monitoring device series, recalled its activity tracker product in 2014 due to continued reports of dizziness, erratic pulse, nausea, pain, and headaches \cite{recall18}. A closer investigation on each model has been reported \cite{Fitbit-radiation}. Some users particularly reported that wearing a Fitbit causes pain in their arm \cite{fitbit_arm}. Relevantly, a recent study suggested that wearing a sleep tracker may worsen an insomnia \cite{nyt19}.

\subsubsection{AirPod}
Unlike Fitbit, a \textit{head-worn} device such as a wireless headset represents another dimension of potential risk. An example is AirPod, a wireless earbud product series by Apple. The product has been found to generate quite high SARs for a Bluetooth device \cite{earphonesAirPods}. Moreover, considering the close proximity to the brain, the health concern around an AirPod becomes more grave than other wearble devices.

\subsubsection{Apple Watch}
The key concern about Apple Watch is the \textit{exposure time}. The smartwatch product performs a wider variety of applications than just health monitoring or music streaming and hence is expected to be worn for a longer time \cite{applebattery}. It has been acknowledged that a longer exposure to EMF radiation can elevate the hazardous impacts over time \cite{weaver06}.

A further problem is that a device still needs to generate radio emission even when not used to interact with a nearby signal source (\textit{i.e.}, access point) \cite{medium}. As such, the time duration of exposure can become longer, which can exacerbate the human EMF exposure problem than one expects.

\subsection{Health Effect}
\subsubsection{Heating}
Temperature elevation of skin is one representative impact on a human body caused by an EMF exposure. The temperature for a skin outer surface normally ranges from 30 to 35$^{\circ}$C. The pain detection threshold temperature for human skin is approximately 43$^{\circ}$C \cite{wu15} and any temperature exceeding it can cause a long-term injury. Heating is considered as a significant impact since it can cause subsequent effects such as cell damage and protein induction \cite{pall18}. It is also known that high-frequency EMF affects the sweat glands (which may serve as helical antennas), peripheral nerves, the eyes and the testes, and may have indirect effects on many organs in the body \cite{book19}.

Recent publications in the field of biology studied health impacts of EMF in frequencies above 6 GHz. In a latest study \cite{kuster19}, EMF power transmitted to the body was analyzed as a function of angle of incidence and polarization, and its relevance to the current guidelines was discussed. Another study \cite{kuster18_bio} determined a maximum averaging area for power density (PD) that limits the maximum temperature increase to a given threshold. Also, considering `bursty' traffic patterns in modern wireless data communications, an analytical approach to `pulsed' heating was developed and applied to assess the peak-to-average temperature ratio as a function of the pulse fraction \cite{kuster18_phys}.

\subsubsection{Neuron}
Also, a recent study reported that exposure of fetal mice to EMF radiation from 800-1900 MHz-rated cellular telephones (24 h/day throughout gestation) affected neuron development and subsequent behavior. Exposed subjects showed memory impairments, as determined by the novel object recognition task \cite{aldad12}.

Furthermore, there is research that proves that these exposure levels can open the blood-brain barrier, leading to neurodegenerative disease and even cancer. Even though FCC states that low levels of radiation emitted by devices like Bluetooth headphones are safe, users are recommended to turn off the Bluetooth function when not in use \cite{appleiphone-7}.

Research has shown that the softer brains of children make them much more susceptible to radiation which means they will absorb more radiation than adults \cite{apple-watch-radiationconcerns}. For this reason, exposure to EMF generated by wearable devices can be particularly dangerous for children. Other factors can affect the specific absorption rate of a device. The more wireless connections, the higher the absorption rate will be. For instance, if one has Bluetooth and Wi-Fi on, the overall SAR will increase.

\subsubsection{Heartbeat}
Disturbance in heartbeats due to EMF exposure has been studied \cite{heartbeat98}. Diurnal rhythms of blood pressure and heart rate of healthy men aged 28-49 years were measured on the basis of data from 24 hours of recordings. The study concluded that exposure to EMF for occupation can result in changes of the diurnal rhythms of blood pressure and heart rate with lowering of their amplitudes and a shift of the acrophase.

\subsection{Current Safety Guidelines}
\subsubsection{Organizations Establishing the Guidelines}
International agencies such as the U.S. Federal Communications Commission (FCC) \cite{fcc_sar}\cite{OET65} and the ICNIRP \cite{ICNIRP74} set the maximum amount of EMF radiation that is allowed to be introduced in the human body without leading to potential health hazards. Similar to other wireless devices, the design of a body-worn device is required to undergo compliances tests based on the safety guidelines. The World Health Organization (WHO), which established the International Electromagnetic Field Project (IEFP), also provides information on health hazards \cite{wireless}.

There are two types of SAR guideline for the general public (head and trunk), which suggests that the subsequent human health impacts depend on the exact area where the device is placed. The SAR limit is 1.6 W/kg averaged over 1 gram (g) of tissue for use against the head and 4.0 W/kg averaged over 10 g of tissue for use on the wrist, based on dosimetric considerations. This limit is recognized in the USA, Canada, and Korea \cite{4197534}. On the other hand, the SAR limit is 2.0 W/kg averaged over 10 g of tissue for use against the head and 4.0 W/kg averaged over 10 g of tissue for use on the wrist, is applied in the EU, Japan, and China \cite{ICNIRP74}\cite{EMC EUROPE}.

\subsubsection{SAR Compliance Test Standards}
The key technical focus is how to determine the SAR measurement methodology. A standard mechanism is required for measurement of the electromagnetic energy absorbed by biological tissue when exposed to radiated electromagnetic energy. These safety tests are conducted with high transmission levels and placed in positions simulating use against the head, with 10 mm separation, and on the wrist, with no separation \cite{rfexposure}. Typically, the mass averaged SAR is computed and compared to the exposure limits set by the regulating standards to prevent heat stress throughout the body and excessive heating of local tissues.

\subsection{Reduction of Human EMF Exposure}\label{sec_stateoftheart_reduction}
Albeit not many, schemes for EMF emission reduction in a wireless system have been proposed in recent literature. Examples include channel precoding \cite{love16} and transmit power control \cite{sambo15}. Note that the human exposure can be reduced if a base station (BS) adopts a power control or adaptive beamforming technique \cite{baracca18}. Also, the exposure level can be reduced when multiple spectrum bands are combined for coordinated use. The reason is that with a higher carrier frequency, a wireless system should reduce the cell size, which leads to more severe threats to human health. Another latest study discovered that the position of the sink is also an important parameter for SAR \cite{Ahmed6}.

\section{SAR Analysis Framework}\label{sec_analysis}
Now one wishes to know how exactly the SAR is evaluated. This section presents mathematical expressions for definition and derivation of SAR averaged over mass and time.

\subsection{Justification of SAR as a Key Metric}
Electromagnetic waves are able to penetrate human tissues and cause oscillating electric fields. Then human tissues absorb these waves, which can affect energy states at the molecular level and thus may lead to harmful effects \cite{pall18}. Specifically, dielectric heating causes a temperature rise in the exposed part of a body \cite{temperature}. This heating is what has been known to result in other potential impacts as a consequence \cite{pall18}.

PD and SAR are the two most widely accepted metrics to measure the intensity and effects of EMF exposure \cite{ieeec95_18}. However, selection of an appropriate metric evaluating the EMF exposure still remains to be an open problem. The FCC suggests PD as a metric measuring the human exposure to EMF generated by devices operating at frequencies higher than 6 GHz \cite{fcc01}. Yet, later, a study suggested that a guideline defined in PD is not efficient to determine the impacts on health issues especially when devices are operating in a very close proximity to the human body such as in an uplink \cite{wu15}.

However, PD cannot evaluate the effect of certain transmission characteristics (\textit{e.g.}, reflection) adequately. Thus, temperature elevation and SAR at a direct contact area are proposed as the appropriate metric for EMF exposure above 6 GHz \cite{temperature}.

For this reason, this paper chooses SAR as a more adequate metric than the skin temperature, which is subject to be dispersed during propagation and be affected by the external atmosphere (\textit{i.e.}, air temperature).

\subsection{Definition of SAR}
The SAR is a quantitative measure of incident energy absorbed per unit of mass and time \cite{Karthik39}. Intuitively, the SAR quantifies the rate at which the human body absorbs energy from an EMF.

Here we review several widely used versions of expressions for SAR. The local SAR value at a point $p$ measured in W/kg \cite{Nasim5G} can be expressed as
\begin{equation}\label{eq_sar_p}
\text{SAR}(p)=\frac{\sigma|E(p)|^{2}}{\rho}{\rm{~~~}}[\text{W/kg}]
\end{equation}
where $\sigma$ is the conductivity of the material in the unit of siemens per meter (S/m); $E$ is the root-mean-square (RMS) of an electric intensity of the body (V/m); and $\rho$ indicates the density of the material (kg/m$^3$).

The SAR can be interpreted with respect to the heat amount as well. Specifically, the SAR can be defined as the ratio of the temperature elevation per unit exposure time. If the heat diffusion is negligibly small during the exposure period, the SAR at an arbitrary point is given by \cite{thermographic}
\begin{equation}\label{eq_sar_t}
\text{SAR} = c\frac{\Delta T}{\Delta t}{\rm{~~~}}[\text{W/kg}]
\end{equation}
where $c$ is the specific heat of the phantom (J/kg/K); $T$ gives the temperature rising at the point of exposure (K); and $t$ is the exposure time length (sec). According to Eq. (\ref{eq_sar_t}), an abrupt temperature elevation due to EMF exposure can be captured by the definition of SAR.

It is noteworthy that the SAR is also inversely proportional to the penetration depth, which suggests a shallower penetration yielding a higher absorption. The SAR at a point on the air-skin boundary can be written as a function of PD$(d,\phi)$, which is given by \cite{chahat_tap12}
\begin{equation}\label{eq_sar_dphi}
\text{SAR}(d,\phi)=\frac{2\text{PD}(d,\phi)(1 - R^{2})}{\delta \rho}\
\end{equation}
where $R$ is the reflection coefficient \cite{wu15_icc}; $\rho$ is the tissue mass density (kg/m$^3$) as defined in Eq. (\ref{eq_sar_p}); and $\delta$ gives the skin penetration depth (m). Notice that PD used in Eq. (\ref{eq_sar_dphi}) is formally written as \cite{love16}
\begin{align}\label{eq_pd_phi}
\text{PD}\left(d, \phi\right) = \frac{P_{t} G_{t}\left(d, \phi\right)}{4 \pi d^2}
\end{align}
where $P_{t}$ is a transmit power; $G_{t}$ is a transmit antenna gain; $d$ is the distance (m) from the transmitter.

Let us elaborate on the angle $\phi$. It indicates an angle formed between the physical orientation of the antenna and the steered angle of departure \cite{love16}. The impact of $\phi$ becomes more dominant with a directional antenna; as such, if an omni-directional antenna is assumed, this parameter can be ignored. This paper adopts the definition of SAR with $\phi$ for the generality of the findings in sequel.

\subsection{Average SAR}
As have been observed in Eqs. (\ref{eq_sar_p}) and (\ref{eq_sar_dphi}), the SAR is defined as the average over a certain amount of mass. Thus, it is referred to as the ``peak spatial average SAR'' in the standard \cite{ieeec95_18}.

However, in most of the current guidelines, a \textit{time average} of the peak spatial average SAR is used as a major metric. For calculation of the time average, a latest guideline set by the IEEE \cite{ieeec95_18} refers to the American National Standards Institute (ANSI) C95.1-1982 \cite{ansi82}. According to the ANSI standard, a set of limits are established as EMF exposure protection guides, which applied to all persons regardless of the nature of the exposure environment. It was stated that for situations involving unrestricted exposure of the body, the protection guides are understood to result in energy deposition averaged over the entire body mass for \textit{any 6-min period} of about 144 joules per kilogram (J/kg) or less. This is equivalent to a SAR of about 0.40 watts per kilogram (W/kg) or less, as specially and temporally averaged over the entire body mass.

The IEEE standard \cite{ieeec95_18} chooses the averaging time of 30 mins for whole body exposure and 6 mins for local exposure. A time averaged exposure calculated over a whole body shall meet the corresponding whole body exposure reference levels (ERLs) regardless of the waveforms, including pulsed fields. According to the standard, the local exposure ERLs are set at 4 times the corresponding whole-body ERL, for frequencies above 2 GHz. (Recall that this paper investigates the EMF exposure at 2.4 and 60 GHz, which is above 2 GHz.)

\subsection{Assumption of Worst-Case Exposure}
It is important to note that this paper considers the \textit{theoretical maximum} exposure that a human user can experience. As introduced in Section \ref{sec_stateoftheart_reduction}, several SAR mitigating techniques have been proposed: \textit{e.g.}, transmit power control \cite{sambo15}\cite{baracca18}. However, this paper does not consider adoption of such mitigation techniques in the SAR definitions presented in Eqs. (\ref{eq_sar_p}) through (\ref{eq_sar_dphi}). It is to guide the consumers towards such a way that the most conservative safety suggestions are provided.

\section{SAR Measurement Methodolgy}\label{sec_measurement}
Now, this section provides technical details on the standard methodology of SAR measurements for compliance tests. Manufacturers of EMF-radiating devices pay serious attention to the tests since any negative test result can require a design change that can negatively impact project cost and schedule \cite{silex}.

\subsection{Measurement Model}
\subsubsection{Complexity of SAR Modeling}
The distribution of SAR in a biological body is complicated since the tissue constitution is differentiated by the body parts \cite{tbme16}. As such, computer simulations are often used to characterize the interaction between an antenna and a load for safety assessment. However, modeling of complex antenna-load structures to match realistic physical conditions using EMF simulations is not straightforward \cite{brish01}\cite{chav03}, and possible discrepancies between simulated and manufactured devices may undermine the accuracy of EMF safety assessment \cite{schmid96}. For this reason, although simulations are still popularly used to conduct initial evaluations \cite{babik05}, the majority of EMF-emitting devices are tested using physical probes \cite{ieeep1528_15}. This setting with a physical probe is also known as a ``conventional SAR measurement'' system \cite{alon15}.

\subsubsection{Time-Average SAR Measurement}
In a conventional SAR measurement system, a liquid-filled model is often used to represent electrical characteristics of body tissues on which a physical electric field probe is applied. Hence, in a conventional measurement setup, the invasiveness of the method, time consumed, and calibration requirements of the probe make the measurement cumbersome and inefficient \cite{Karthik39}.

\subsection{Current Measurement Methodology}
Currently, the most popular measurement methodology to obtain official SAR measurements is the dosimetric assessment system (DASY) \cite{speag.swis}.

\subsubsection{Setup of DASY}
\begin{figure}
\centering
\begin{subfigure}[b]{\linewidth}
\centering
\includegraphics[width = \linewidth]{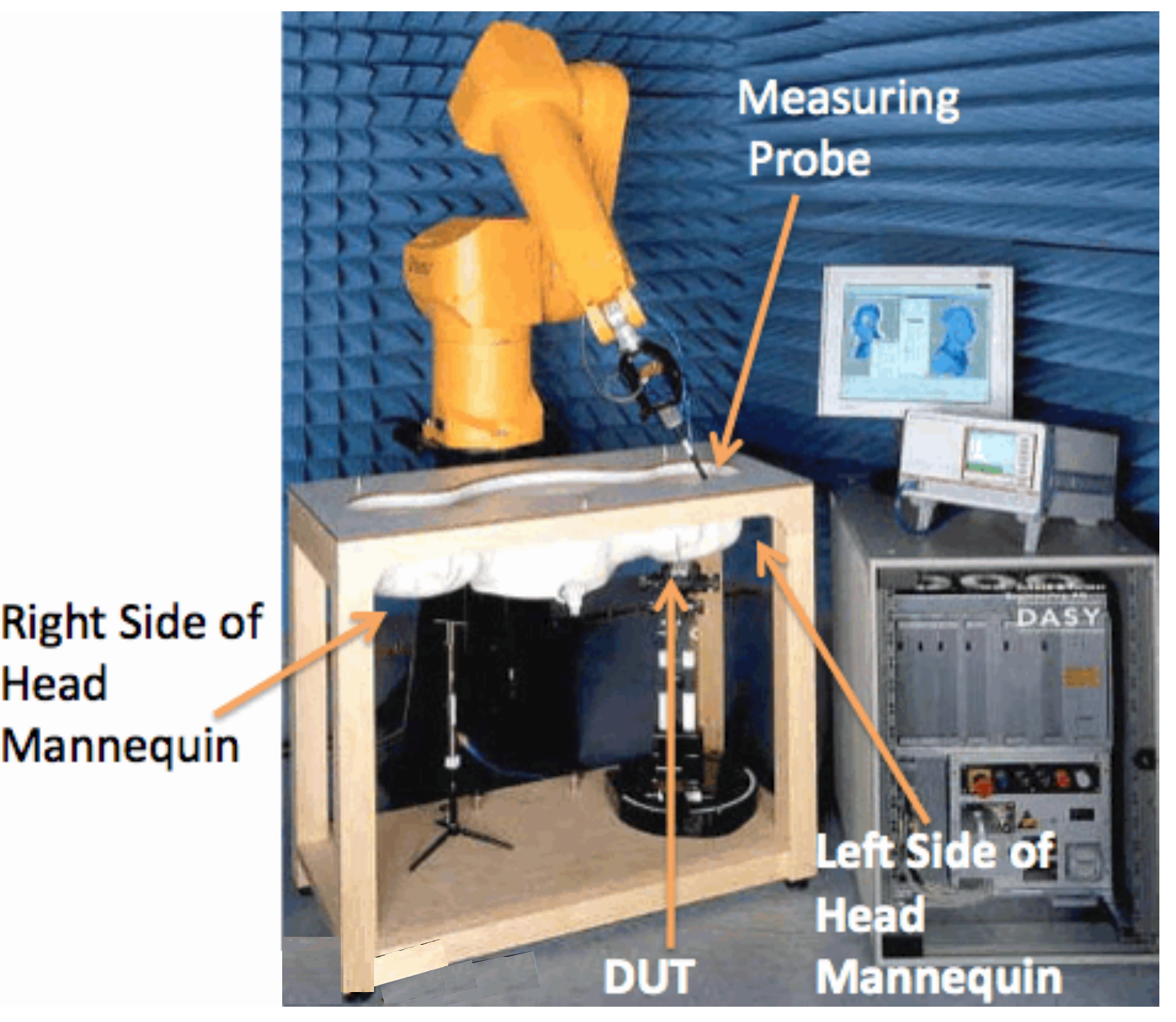}
\caption{An example setup \cite{Wang2}}
\label{fig_dasy_setup}
\end{subfigure}\hfill
\vspace{0.2 in}
\begin{subfigure}[b]{\linewidth}
\centering
\includegraphics[width = \linewidth]{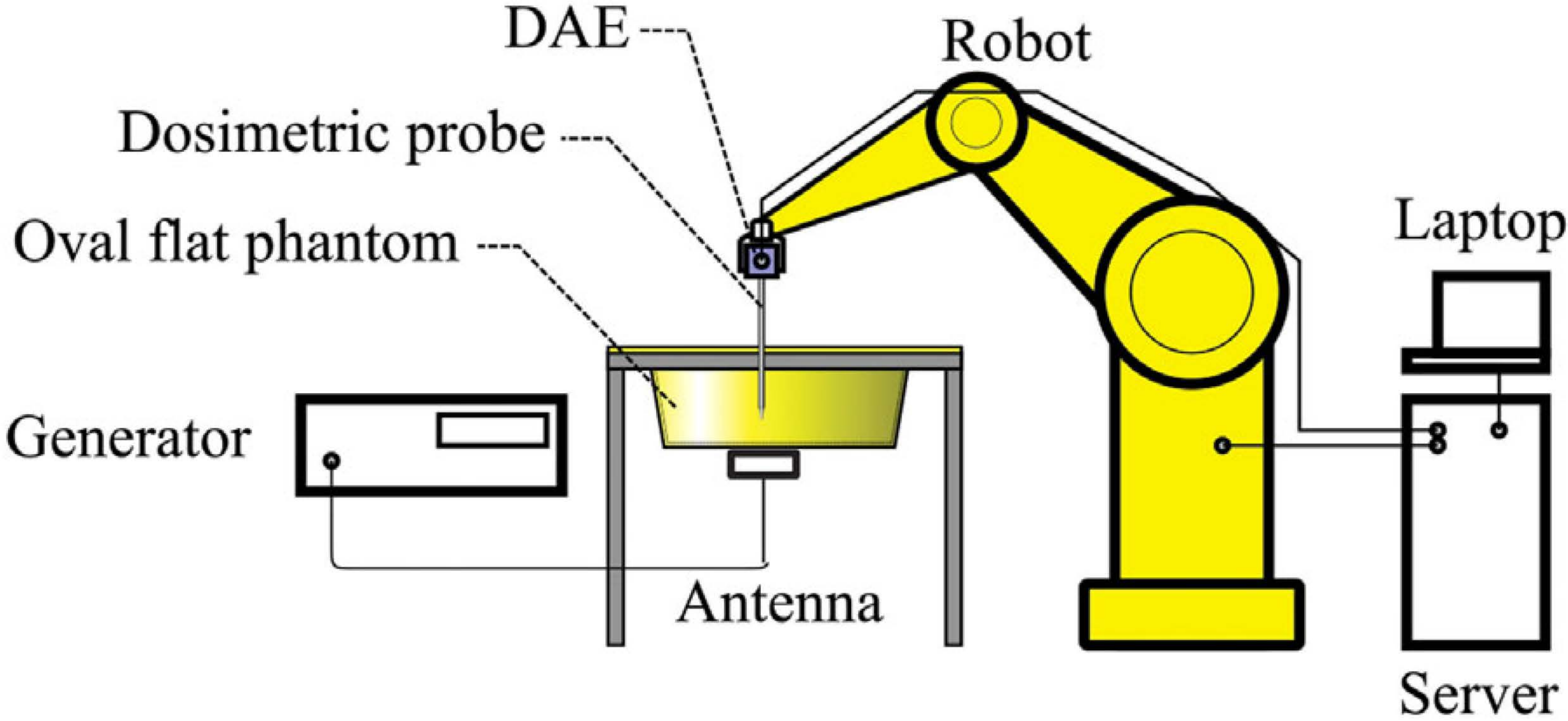}
\caption{Schematic \cite{Thielens}}
\label{fig_dasy_schematic}
\end{subfigure}
\caption{A typical setup of DASY}
\label{fig_dasy}
\end{figure}

Fig. \ref{fig_dasy} shows how a SAR measurement can be set up with the DASY. As illustrated in Fig. \ref{fig_dasy_setup}, a DASY system is made up of a high precision robot (the yellow-colored robot as also shown in Fig. \ref{fig_dasy_schematic}), a probe alignment sensor, a phantom, a robot controller, a controlled measurement server, a data acquisition electronics (DAE), and a probe. The robot includes 6 axes that can move to the precision position defined by the DASY software. The DASY software can define the area that is detected by the probe. The SAR measurement is conducted with the dosimetric probe, which is specially designed and calibrated for use in liquid, with high permittivity. The dosimetric probe has a special calibration in liquid at different frequencies.

\subsubsection{Measurement of Time-Averaged SAR}
In a DASY, a time-averaged SAR is calculated by specifying the device dimensions first in the project setup \cite{speag.swis}. Next, the phantom section, test distance, the position of the device under test (DUT), and communications system are specified. The area scan and zoom scan are performed accordingly. The duration of the scan is specified in the scan duration field. The default value is 6 minutes, compliant with the current safety guideline \cite{ieeec95_18}.

A typical measurement procedure can be summarized as follows \cite{ieeep1528_15}:
\begin{itemize}
\item \textit{SAR Reference Measurement:} Before a time-average SAR test, a local SAR is measured at a stationary reference point where the SAR exceeds the lower detection limit of the measurement system.
\item \textit{Area Scan:} Once a reference is found, it is necessary to determine peak SAR positions. An electric field probe moves through the tissue-equivalent liquid in a specific anthropomorphic mannequin (SAM) or a flat phantom to find the approximate location of the SAR peak. The distance between the phantom and the probe has to be more than half the probe diameter; otherwise, the peak SAR is measured at a higher value than actual and in turn the uncertainty is increased. The measured values are interpolated to determine peak locations. The rationale of the interpolation is that a sensor behind a probe tip is usually not able to measure a local peak SAR occurring at the surface of homogeneous phantoms \cite{Wang2}.
\item \textit{Zoom Scan:} The goal of a zoom scan is to determine cube averaged SAR. Zoom scans surrounding one or more of these peak locations are performed to finally determine the value of a peak spatial-average SAR.
\end{itemize}

\subsubsection{Controversy}
In the design of a SAR measurement method, the main challenge is how to efficiently mock the human body in a test setup. The reason is that it is difficult to account for the numerous critical variables affecting the test responses--\textit{i.e.}, the subject's age, the amount of subcutaneous fat, the physical condition of the individual, etc \cite{shel99}.

This complexity has been noted recently in practice. It is claimed that the current SAR measurement methodology is not general enough to ``represent'' the diverse characteristics of a human body \cite{earphonesAirPods}. The methodology was developed in 1989 and the dummy head, which is used for the measurements, is equivalent to that of a 6 feet 2-inch person weighing 220 pounds. The human brain is represented by a simple mixture of water and electrolytes, and the massive head is subjected to only 6 minutes of peak activity before taking the measurements. Although these are only some of the issues of the test, they are enough to show that it may not represent an average person's usage in 2019.

The similar inefficiency occurs when a measurement is conducted using animal subjects. Previous studies have produced conflicting evidence regarding the health effects of exposure to EMF due to methodological differences in experimental setups, exposure durations, the animal models used, and the behavioral test(s) used to asses learning and/or memory \cite{wang17}.

\subsection{Recent Proposal on EMF Exposure Measurement Methodology}
In order to address the aforementioned issues on the current SAR measurement methodology, recent studies have proposed alternative approaches to assess the human EMF exposure.

\subsubsection{Temperature Measurement}
In addition to the electric field probe dosimetry systems, a temperature-based dosimetry system was proposed using an array of optical fiber thermal sensors positioned inside a phantom, showing good agreement with electric field probe measurements \cite{alon15_13}. However, spatial resolution has been limited because of the large number of temperature probes that need to be positioned inside the phantom invasively.

Recently, it has been shown that magnetic resonance imaging (MRI) temperature mapping has been used to quantify EMF energy deposition induced by MRI coils \cite{alon15_14}-\cite{alon15_16} and other antennas \cite{alon15_17}\cite{alon15_18}, where both radio frequency (RF) heating and temperature mapping were conducted inside the bore of the MRI scanner.

\subsubsection{Chronic Exposure}
The National Toxicology Program (NTP) \cite{capstick17_7} of the National Institute of Environmental Health Sciences (NIEHS) defines a cancer bioassay as ``exposure of laboratory animals to a chemical, biological, or physical agent for at least 2 years.'' The NTP test also requires that the exposure level be practically constant over the lifetime of the subjects and that the animals be free to move within the individual cages. A recent work \cite{capstick17} presented results of tests in a room-size exposure chamber capable of housing a large number of rodents for an NTP assessment of the toxicity and carcinogenicity of RF signals emitted from widely used cell phones.

\begin{table}[t]
\centering
\caption{SAR and key parameters for AirPod, Fitbit and Snap wearable video camera \cite{RF-Exposure-Info-4246043}-\cite{fcc.report-3974904}}
\label{table_list1}
\begin{tabular}{|c|c|c|c|}
\hline 
\textbf{\tiny{}Wearable Device}{\cellcolor{gray!20}} & \textbf{\tiny{}Parameter}{\cellcolor{gray!20}} & {\tiny{}802.11b}{\cellcolor{gray!20}} & {\tiny{}Bluetooth }{\cellcolor{gray!20}}\tabularnewline
\hline 
& {\tiny{}Frequency } & {\tiny{}N/A } & {\tiny{}2441 }\tabularnewline
\cline{2-4} 
& {\tiny{}Spacing (mm) } & {\tiny{}N/A } & {\tiny{}10 }\tabularnewline
\cline{2-4} 
{\tiny{}Apple AirPod A2031,} & {\tiny{}Conducted Power {[}dBm{]} } & {\tiny{}N/A } & {\tiny{}12.5 }\tabularnewline
\cline{2-4} 
{\tiny{}Left ear (next to mouth)} & {\tiny{}Duty Cycle (percentage) } & {\tiny{}N/A } & {\tiny{}100 }\tabularnewline
\cline{2-4} 
& {\tiny{}Reported SAR (averaged over 1 g) {[}W/Kg{]} } & {\tiny{}N/A } & {\tiny{}0.071 }\tabularnewline
\hline 
& {\tiny{}Frequency } & {\tiny{}N/A } & {\tiny{}2441 }\tabularnewline
\cline{2-4} 
& {\tiny{}Spacing (mm) } & {\tiny{}N/A } & {\tiny{}0 }\tabularnewline
\cline{2-4} 
{\tiny{}Apple AirPod A2031,} & {\tiny{}Conducted Power {[}dBm{]} } & {\tiny{}N/A } & {\tiny{}12.5 }\tabularnewline
\cline{2-4} 
{\tiny{}Left ear (body-mounted)} & {\tiny{}Duty Cycle (percentage) } & {\tiny{}N/A } & {\tiny{}100 }\tabularnewline
\cline{2-4} 
& {\tiny{}Reported SAR (averaged over 10 g) {[}W/Kg{]} } & {\tiny{}N/A } & {\tiny{}0.501 }\tabularnewline
\hline 
& {\tiny{}Frequency } & {\tiny{}N/A } & {\tiny{}2441 }\tabularnewline
\cline{2-4} 
& {\tiny{}Spacing (mm) } & {\tiny{}N/A } & {\tiny{}10 }\tabularnewline
\cline{2-4} 
{\tiny{}Apple AirPod A2032,} & {\tiny{}Conducted Power {[}dBm{]} } & {\tiny{}N/A } & {\tiny{}12.5 }\tabularnewline
\cline{2-4} 
{\tiny{}Right ear (next to mouth)}& {\tiny{}Duty Cycle (percentage) } & {\tiny{}N/A } & {\tiny{}100 }\tabularnewline
\cline{2-4} 
& {\tiny{}Reported SAR (averaged over 1 g) {[}W/Kg{]} } & {\tiny{}N/A } & {\tiny{}0.095 }\tabularnewline
\hline 
& {\tiny{}Frequency } & {\tiny{}N/A } & {\tiny{}2441 }\tabularnewline
\cline{2-4} 
& {\tiny{}Spacing (mm) } & {\tiny{}N/A } & {\tiny{}0 }\tabularnewline
\cline{2-4} 
{\tiny{}Apple AirPod A2032,} & {\tiny{}Conducted Power {[}dBm{]} } & {\tiny{}N/A } & {\tiny{}12.5 }\tabularnewline
\cline{2-4} 
{\tiny{}Right ear (body-mounted)} & {\tiny{}Duty Cycle (percentage) } & {\tiny{}N/A } & {\tiny{}100 }\tabularnewline
\cline{2-4} 
& {\tiny{}Reported SAR (averaged over 10 g) {[}W/Kg{]} } & {\tiny{}N/A } & {\tiny{}0.581 }\tabularnewline
\hline 
& {\tiny{}Frequency } & {\tiny{}N/A } & {\tiny{}2441 }\tabularnewline
\cline{2-4} 
& {\tiny{}Spacing (mm) } & {\tiny{}N/A } & {\tiny{}0 }\tabularnewline
\cline{2-4} 
{\tiny{}Fitbit xRAFB202 } & {\tiny{}Conducted Power {[}dBm{]} } & {\tiny{}18.3 } & {\tiny{}N/A }\tabularnewline
\cline{2-4} 
& {\tiny{}Duty Cycle (percentage) } & {\tiny{}100 } & {\tiny{}N/A }\tabularnewline
\cline{2-4} 
& {\tiny{}Reported SAR (averaged over 10 g) {[}W/Kg{]} } & {\tiny{}1.124 } & {\tiny{}N/A }\tabularnewline
\hline 
& {\tiny{}Frequency } & {\tiny{}2437 } & {\tiny{}N/A }\tabularnewline
\cline{2-4} 
& {\tiny{}Spacing (mm) } & {\tiny{}0 } & {\tiny{}N/A }\tabularnewline
\cline{2-4} 
{\tiny{}Fitbit xRAFB503 } & {\tiny{}Conducted Power {[}dBm{]} } & {\tiny{}18.4 } & {\tiny{}N/A }\tabularnewline
\cline{2-4} 
& {\tiny{}Duty Cycle (percentage) } & {\tiny{}100 } & {\tiny{}N/A }\tabularnewline
\cline{2-4} 
& {\tiny{}Reported SAR (averaged over 10 g) {[}W/Kg{]} } & {\tiny{}0.45 } & {\tiny{}N/A }\tabularnewline
\hline 
& {\tiny{}Frequency } & {\tiny{}2412 } & {\tiny{}N/A }\tabularnewline
\cline{2-4} 
& {\tiny{}Spacing (mm) } & {\tiny{}0 } & {\tiny{}N/A }\tabularnewline
\cline{2-4} 
{\tiny{}Snap wearable video camera,} & {\tiny{}Conducted Power {[}dBm{]} } & {\tiny{}12.65 } & {\tiny{}N/A }\tabularnewline
\cline{2-4} 
{\tiny{}2AIRN-002 Veronica} & {\tiny{}Duty Cycle (percentage) } & {\tiny{}92-96 } & {\tiny{}N/A }\tabularnewline
\cline{2-4} 
& {\tiny{}Reported SAR (averaged over 10 g) {[}W/Kg{]} } & {\tiny{}0.94 } & {\tiny{}N/A }\tabularnewline
\hline 
& {\tiny{}Frequency } & {\tiny{}2412 } & {\tiny{}N/A }\tabularnewline
\cline{2-4} 
& {\tiny{}Spacing (mm) } & {\tiny{}0 } & {\tiny{}N/A }\tabularnewline
\cline{2-4} 
{\tiny{}Snap wearable video camera,} & {\tiny{}Conducted Power {[}dBm{]} } & {\tiny{}12.65 } & {\tiny{}N/A }\tabularnewline
\cline{2-4} 
{\tiny{}2AIRN-002 Nico} & {\tiny{}Duty Cycle (percentage) } & {\tiny{}92-96 } & {\tiny{}N/A }\tabularnewline
\cline{2-4} 
& {\tiny{}Reported SAR (averaged over 10 g) {[}W/Kg{]} } & {\tiny{}1 } & {\tiny{}N/A }\tabularnewline
\hline 
\end{tabular}
\end{table}

\begin{center}
\begin{table*}[tbh]
\caption{SAR and key parameters for Apple Watch \cite{RF-Exposure-Info-1-3549293}-\cite{RF-Exposure-Info-3555643}}
\label{table_list2}
\begin{centering}
\centering{}%
\begin{tabular}{|c|c|c|c|c|c|c|c|}
\hline 
\textbf{\tiny{}Wearable Device}{\cellcolor{gray!20}} & \textbf{\tiny{}Parameter}{\cellcolor{gray!20}} & {\tiny{}UMTS 850}{\cellcolor{gray!20}} & {\tiny{}UMTS 1750}{\cellcolor{gray!20}} & {\tiny{}LTE Band 7}{\cellcolor{gray!20}} & {\tiny{}LTE Band 26}{\cellcolor{gray!20}} & {\tiny{}802.11b}{\cellcolor{gray!20}} & {\tiny{}Bluetooth}{\cellcolor{gray!20}}\tabularnewline
\hline 
& {\tiny{}Frequency} & {\tiny{}836.6} & {\tiny{}1732.4} & {\tiny{}N/A} & {\tiny{}844} & {\tiny{}2437} & {\tiny{}2441}\tabularnewline
\cline{2-8} 
& {\tiny{}Spacing (mm)} & {\tiny{}10} & {\tiny{}10} & {\tiny{}N/A} & {\tiny{}10} & {\tiny{}10} & {\tiny{}10}\tabularnewline
\cline{2-8} 
& {\tiny{}Housing Type} & {\tiny{}Stainless Steel} & {\tiny{}Ceramic} & {\tiny{}N/A} & {\tiny{}Stainless Steel} & {\tiny{}Aluminum} & {\tiny{}Ceramic}\tabularnewline
\cline{2-8} 
{\tiny{}Apple Watch A1860 (next to mouth) } & {\tiny{}Wrist Band Type} & {\tiny{}Sport} & {\tiny{}Sport} & {\tiny{}N/A} & {\tiny{}Metal Links} & {\tiny{}Sport} & {\tiny{}Sport}\tabularnewline
\cline{2-8} 
& {\tiny{}Conducted Power {[}dBm{]}} & {\tiny{}22.89} & {\tiny{}23.43} & {\tiny{}N/A} & {\tiny{}22.8} & {\tiny{}19.47} & {\tiny{}12.98}\tabularnewline
\cline{2-8} 
& {\tiny{}Duty Cycle (percentage)} & {\tiny{}100} & {\tiny{}100} & {\tiny{}N/A} & {\tiny{}100} & {\tiny{}98.2} & {\tiny{}100}\tabularnewline
\cline{2-8} 
& {\tiny{}Reported SAR (averaged over 1 g) {[}W/Kg{]}} & {\tiny{}0.112} & {\tiny{}0.526} & {\tiny{}N/A} & {\tiny{}0.109} & {\tiny{}0.089} & {\tiny{}0.094}\tabularnewline
\hline 
& {\tiny{}Frequency} & {\tiny{}836.6} & {\tiny{}1732.4} & {\tiny{}N/A} & {\tiny{}844} & {\tiny{}2437} & {\tiny{}2441}\tabularnewline
\cline{2-8} 
& {\tiny{}Spacing (mm)} & {\tiny{}0} & {\tiny{}0} & {\tiny{}N/A} & {\tiny{}0} & {\tiny{}0} & {\tiny{}0}\tabularnewline
\cline{2-8} 
& {\tiny{}Housing Type} & {\tiny{}Ceramic} & {\tiny{}Ceramic} & {\tiny{}N/A} & {\tiny{}Ceramic} & {\tiny{}Aluminum} & {\tiny{}Aluminum}\tabularnewline
\cline{2-8} 
{\tiny{}Apple Watch A1860 (body-mounted) } & {\tiny{}Wrist Band Type} & {\tiny{}Sport} & {\tiny{}Metal Links} & {\tiny{}N/A} & {\tiny{}Metal Loop} & {\tiny{}Sport} & {\tiny{}Sport}\tabularnewline
\cline{2-8} 
& {\tiny{}Conducted Power {[}dBm{]}} & {\tiny{}22.89} & {\tiny{}23.43} & {\tiny{}N/A} & {\tiny{}22.8} & {\tiny{}19.47} & {\tiny{}12.98}\tabularnewline
\cline{2-8} 
& {\tiny{}Duty Cycle (percentage)} & {\tiny{}100} & {\tiny{}100} & {\tiny{}N/A} & {\tiny{}100} & {\tiny{}98.2} & {\tiny{}100}\tabularnewline
\cline{2-8} 
& {\tiny{}Reported SAR (averaged over 10 g) {[}W/Kg{]}} & {\tiny{}0.026} & {\tiny{}0.179} & {\tiny{}N/A} & {\tiny{}0.024} & {\tiny{}0.029} & {\tiny{}0.034}\tabularnewline
\hline 
& {\tiny{}Frequency} & {\tiny{}836.6} & {\tiny{}1732.4} & {\tiny{}N/A} & {\tiny{}819} & {\tiny{}2437} & {\tiny{}2441}\tabularnewline
\cline{2-8} 
& {\tiny{}Spacing (mm)} & {\tiny{}10} & {\tiny{}10} & {\tiny{}N/A} & {\tiny{}10} & {\tiny{}10} & {\tiny{}10}\tabularnewline
\cline{2-8} 
& {\tiny{}Housing Type} & {\tiny{}Stainless Steel} & {\tiny{}Ceramic} & {\tiny{}N/A} & {\tiny{}Stainless Steel} & {\tiny{}Aluminum} & {\tiny{}Aluminum}\tabularnewline
\cline{2-8} 
{\tiny{}Apple Watch A1861 (next to mouth)} & {\tiny{}Wrist Band Type} & {\tiny{}Sport} & {\tiny{}Sport} & {\tiny{}N/A} & {\tiny{}Metal Links} & {\tiny{}Sport} & {\tiny{}Sport}\tabularnewline
\cline{2-8} 
& {\tiny{}Conducted Power {[}dBm{]}} & {\tiny{}22.89 } & {\tiny{}23.43} & {\tiny{}N/A} & {\tiny{}22.8 } & {\tiny{}19.47} & {\tiny{}12.98 }\tabularnewline
\cline{2-8} 
& {\tiny{}Duty Cycle (percentage)} & {\tiny{}100} & {\tiny{}100} & {\tiny{}N/A} & {\tiny{}100} & {\tiny{}98.2} & {\tiny{}100}\tabularnewline
\cline{2-8} 
& {\tiny{}Reported SAR (averaged over 1 g) {[}W/Kg{]}} & {\tiny{}0.112} & {\tiny{}0.526} & {\tiny{}N/A} & {\tiny{}0.109} & {\tiny{}0.166} & {\tiny{}0.130}\tabularnewline
\hline 
& {\tiny{}Frequency} & {\tiny{}836.6} & {\tiny{}1732.4} & {\tiny{}N/A} & {\tiny{}819} & {\tiny{}2437} & {\tiny{}2441}\tabularnewline
\cline{2-8} 
& {\tiny{}Spacing (mm)} & {\tiny{}0} & {\tiny{}0} & {\tiny{}N/A} & {\tiny{}0} & {\tiny{}0} & {\tiny{}0}\tabularnewline
\cline{2-8} 
& {\tiny{}Housing Type} & {\tiny{}Ceramic} & {\tiny{}Ceramic} & {\tiny{}N/A} & {\tiny{}Ceramic} & {\tiny{}Aluminum} & {\tiny{}Aluminum}\tabularnewline
\cline{2-8} 
{\tiny{}Apple Watch A1861 (body-mounted)} & {\tiny{}Wrist Band Type} & {\tiny{}Sport} & {\tiny{}Metal Loop} & {\tiny{}N/A} & {\tiny{}Metal Loop} & {\tiny{}Sport} & {\tiny{}Sport}\tabularnewline
\cline{2-8} 
& {\tiny{}Conducted Power {[}dBm{]}} & {\tiny{}23} & {\tiny{}23.5} & {\tiny{}N/A} & {\tiny{}22.88} & {\tiny{}19.47} & {\tiny{}12.98}\tabularnewline
\cline{2-8} 
& {\tiny{}Duty Cycle (percentage)} & {\tiny{}100} & {\tiny{}100} & {\tiny{}N/A} & {\tiny{}100} & {\tiny{}98.2} & {\tiny{}100}\tabularnewline
\cline{2-8} 
& {\tiny{}Reported SAR (averaged over 10 g) {[}W/Kg{]}} & {\tiny{}0.03} & {\tiny{}0.344} & {\tiny{}N/A} & {\tiny{}0.018} & {\tiny{}0.083} & {\tiny{}0.070}\tabularnewline
\hline 
& {\tiny{}Frequency} & {\tiny{}826.4} & {\tiny{}N/A} & {\tiny{}2560} & {\tiny{}819} & {\tiny{}2437} & {\tiny{}2441}\tabularnewline
\cline{2-8} 
& {\tiny{}Spacing (mm)} & {\tiny{}10} & {\tiny{}N/A} & {\tiny{}10} & {\tiny{}10} & {\tiny{}10} & {\tiny{}10}\tabularnewline
\cline{2-8} 
& {\tiny{}Housing Type} & {\tiny{}Stainless Steel} & {\tiny{}N/A} & {\tiny{}Aluminum} & {\tiny{}Stainless Steel} & {\tiny{}Aluminum} & {\tiny{}Stainless Steel}\tabularnewline
\cline{2-8} 
{\tiny{}Apple Watch A1889 (next to mouth)} & {\tiny{}Wrist Band Type} & {\tiny{}Metal Loop} & {\tiny{}N/A} & {\tiny{}Sport} & {\tiny{}Metal Links} & {\tiny{}Sport} & {\tiny{}Sport}\tabularnewline
\cline{2-8} 
& {\tiny{}Conducted Power {[}dBm{]}} & {\tiny{}23.39} & {\tiny{}N/A} & {\tiny{}22.9} & {\tiny{}22.52} & {\tiny{}19.49} & {\tiny{}12.81}\tabularnewline
\cline{2-8} 
& {\tiny{}Duty Cycle (percentage)} & {\tiny{}100} & {\tiny{}N/A} & {\tiny{}100} & {\tiny{}100} & {\tiny{}98.2} & {\tiny{}100}\tabularnewline
\cline{2-8} 
& {\tiny{}Reported SAR (averaged over 1 g) {[}W/Kg{]}} & {\tiny{}0.076} & {\tiny{}N/A} & {\tiny{}0.29} & {\tiny{}0.1} & {\tiny{}0.109} & {\tiny{}0.107}\tabularnewline
\hline 
& {\tiny{}Frequency} & {\tiny{}826.4} & {\tiny{}N/A} & {\tiny{}2560} & {\tiny{}819} & {\tiny{}2437} & {\tiny{}2441}\tabularnewline
\cline{2-8} 
& {\tiny{}Spacing (mm)} & {\tiny{}0} & {\tiny{}N/A} & {\tiny{}0} & {\tiny{}0} & {\tiny{}0} & {\tiny{}0}\tabularnewline
\cline{2-8} 
& {\tiny{}Housing Type} & {\tiny{}Ceramic} & {\tiny{}N/A} & {\tiny{}Aluminum} & {\tiny{}Ceramic} & {\tiny{}Aluminum} & {\tiny{}Aluminum}\tabularnewline
\cline{2-8} 
{\tiny{}Apple Watch A1889 (body-mounted)} & {\tiny{}Wrist Band Type} & {\tiny{}Sport} & {\tiny{}N/A} & {\tiny{}Sport} & {\tiny{}Sport} & {\tiny{}Sport} & {\tiny{}Sport}\tabularnewline
\cline{2-8} 
& {\tiny{}Conducted Power {[}dBm{]}} & {\tiny{}23.39} & {\tiny{}N/A} & {\tiny{}22.9} & {\tiny{}22.52} & {\tiny{}19.49} & {\tiny{}12.81}\tabularnewline
\cline{2-8} 
& {\tiny{}Duty Cycle (percentage)} & {\tiny{}100} & {\tiny{}N/A} & {\tiny{}100} & {\tiny{}100} & {\tiny{}98.2} & {\tiny{}100}\tabularnewline
\cline{2-8} 
& {\tiny{}Reported SAR (averaged over 10 g) {[}W/Kg{]}} & {\tiny{}0.023} & {\tiny{}N/A} & {\tiny{}0.146} & {\tiny{}0.024} & {\tiny{}0.036 } & {\tiny{}0.033}\tabularnewline
\hline 
& {\tiny{}Frequency} & {\tiny{}836.6} & {\tiny{}N/A} & {\tiny{}2510} & {\tiny{}819} & {\tiny{}2437} & {\tiny{}2441}\tabularnewline
\cline{2-8} 
& {\tiny{}Spacing (mm)} & {\tiny{}10} & {\tiny{}N/A} & {\tiny{}10} & {\tiny{}10} & {\tiny{}10} & {\tiny{}10}\tabularnewline
\cline{2-8} 
& {\tiny{}Housing Type} & {\tiny{}Stainless Steel} & {\tiny{}N/A} & {\tiny{}Stainless Steel} & {\tiny{}Ceramic} & {\tiny{}Aluminum} & {\tiny{}Aluminum}\tabularnewline
\cline{2-8} 
{\tiny{}Apple Watch A1891 (next to mouth)} & {\tiny{}Wrist Band Type} & {\tiny{}Metal Links} & {\tiny{}N/A} & {\tiny{}Sport} & {\tiny{}Metal Links} & {\tiny{}Sport} & {\tiny{}Sport}\tabularnewline
\cline{2-8} 
& {\tiny{}Conducted Power {[}dBm{]}} & {\tiny{}23.11} & {\tiny{}N/A} & {\tiny{}23.04} & {\tiny{}21.87} & {\tiny{}19.49} & {\tiny{}12.81}\tabularnewline
\cline{2-8} 
& {\tiny{}Duty Cycle (percentage)} & {\tiny{}100} & {\tiny{}N/A} & {\tiny{}100} & {\tiny{}100} & {\tiny{}100} & {\tiny{}100}\tabularnewline
\cline{2-8} 
& {\tiny{}Reported SAR (averaged over 1 g) {[}W/Kg{]}} & {\tiny{}0.134} & {\tiny{}N/A} & {\tiny{}0.354} & {\tiny{}0.11} & {\tiny{}0.144} & {\tiny{}0.176}\tabularnewline
\hline 
& {\tiny{}Frequency} & {\tiny{}836.6} & {\tiny{}N/A} & {\tiny{}2535} & {\tiny{}819} & {\tiny{}2437} & {\tiny{}2441}\tabularnewline
\cline{2-8} 
& {\tiny{}Spacing (mm)} & {\tiny{}0} & {\tiny{}N/A} & {\tiny{}0} & {\tiny{}0} & {\tiny{}0} & {\tiny{}0}\tabularnewline
\cline{2-8} 
& {\tiny{}Housing Type} & {\tiny{}Ceramic} & {\tiny{}N/A} & {\tiny{}Ceramic} & {\tiny{}Ceramic} & {\tiny{}Aluminum} & {\tiny{}Aluminum}\tabularnewline
\cline{2-8} 
{\tiny{}Apple Watch A1891 (body-mounted)} & {\tiny{}Wrist Band Type} & {\tiny{}Sport} & {\tiny{}N/A} & {\tiny{}Sport} & {\tiny{}Metal Loop} & {\tiny{}Sport} & {\tiny{}Sport}\tabularnewline
\cline{2-8} 
& {\tiny{}Conducted Power {[}dBm{]}} & {\tiny{}23.11} & {\tiny{}N/A} & {\tiny{}22.9} & {\tiny{}22.83} & {\tiny{}19.49} & {\tiny{}12.81}\tabularnewline
\cline{2-8} 
& {\tiny{}Duty Cycle (percentage)} & {\tiny{}100} & {\tiny{}N/A} & {\tiny{}100} & {\tiny{}100} & {\tiny{}100} & {\tiny{}100}\tabularnewline
\cline{2-8} 
& {\tiny{}Reported SAR (averaged over 10 g) {[}W/Kg{]}} & {\tiny{}0.028} & {\tiny{}N/A} & {\tiny{}0.178} & {\tiny{}0.021} & {\tiny{}0.085} & {\tiny{}0.102}\tabularnewline
\hline 
\end{tabular}
\par\end{centering}
\end{table*}
\end{center}

\subsubsection{Limitations}
The major limitation of the MRI-based temperature measurement is the compatibility of the commercial wireless devices with the MRI. Motivated from the incompatibility, a recent work proposed a seminal framework integrating an MRI-compatible dipole antenna and a non–MRI-compatible mobile phone via phantom temperature change measurements \cite{alon15}. This MRI-based measurement technique has lately been extended to the mmW devices \cite{alon17}.

The chronic exposure studies showed limitations in the sense that they relied on animals. It has been reported that animals provide accurate approximation in some parts. For instance, there are structural similarities and comparable dimensions of the eyes between rabbits and humans, which makes possible the use of the same power density threshold \cite{gwen93}. Nevertheless, it is usually complicated to extrapolate scientific findings using animals to humans due to different biochemical characteristics from humans \cite{brack08}. 

The key discrepancy in the result of SAR measurements have been discussed in detail \cite{humanrat18}. The peak temperature elevation in the human brain was lower than that in the rat model, mainly because of difference in depth from the scalp.

Moreover, the thermal damage is differentiated between rats and humans. The damage depends on tissue sensitivity, temperature, and exposure time. The cumulative equivalent minutes at 43$^{0}$C are used as a model to calculate the thermal dose \cite{humanrat18_24}. In this evaluation, the thermal time constant is essential. Focusing on the brain, which is a highly heat-sensitive tissue, the thermal time constant in humans was more than twice that in rats. Thus, the exposure time required for thermal damage is correspondingly increased. Furthermore, because the characteristics of the temperature rise of deep tissues in rats and humans are different, extrapolation from small animals to humans in deep tissues needs further attention.

As such, this paper suggests that the improvement in the SAR measurement methodology takes place in the line of improving the aforementioned two technical limitations.

\section{Analysis of SAR Data from Commercial Wearable Communications Devices}\label{sec_devices}
We understand the widespread curiosity on the safety of using the commercial wearable communications devices. Motivated from the interest, we provide an extensive investigation on the SAR levels of the wearable communications devices that are in the market, which distinguishes this work from prior surveys \cite{easychair}\cite{traj16}\cite{elsevier18}. The SAR values that are discussed in this section are either (i) revealed in each product's data sheet or (ii) in case not disclosed, calculated based on the analysis and measurement methods discussed in Sections \ref{sec_analysis} and \ref{sec_measurement}, respectively.

\subsection{Apple AirPod, Fitbit, and Snap Wearable Video Camera}
Table \ref{table_list1} lists the SARs for Apple AirPod, Fitbit, and Snap wearable video camera. Notice that for accurate understanding, along with the SAR, other technical details are also provided--\textit{i.e.}, the carrier frequency, physical spacing from the skin, conducted power, and duty cycle of a signal.

It is also noteworthy that the amount of mass for calculation of a SAR differs according to the body part on which the device is worn--\textit{i.e.}, averaged over 1 g and 10 g for head and body, respectively. For instance, the first two rows of Table \ref{table_list1} indicate Apple AirPod A2031. The SAR averaging mass for SAR evaluation is 1 g and 10 g when worn (i) next to mouth and (ii) on body, respectfully.

Devices that are designed to be worn on the wrist may operate in speaker mode for voice communication, with the device worn on the wrist and positioned next to the mouth. When next-to-mouth SAR evaluation is required, the device is positioned at 10 mm from a flat phantom filled with head tissue-equivalent medium. The device is evaluated with wrist bands strapped together to represent normal use conditions. Devices that are designed or intended for use on extremities or mainly operated in extremity only exposure conditions; \textit{i.e.}, hands, wrists, feet, and ankles, may require extremity SAR evaluation. In this case, the device is evaluated with the back of the device touching the flat phantom, which is filled with a body tissue-equivalent medium.

The devices that are listed in Table \ref{table_list1} operate based on two wireless technologies--namely, IEEE 802.11b and Bluetooth. The carrier frequency is a key factor that dominates the level of SAR \cite{milcom19}. The rationale is inferred from Eq. (\ref{eq_sar_dphi}). A wave with a higher frequency penetrates shallower to the human skin, which in turn lowers $\delta$ and hence increases the SAR. Since 802.11b and Bluetooth operate at the same frequency of 2.4 GHz, we defer the discussion on the impact of carrier frequency for now.

AirPod shows higher SARs for operating based on Bluetooth. According to Table \ref{table_list1}, the SAR for AirPods is 0.581 W/kg for the left earbud and 0.501 W/kg for the right ear. That makes for a combined 1.082 W/kg when worn in both ears. In comparison, the SAR for an iPhone XS is 1.19 W/kg, or just 10\% more than that of the AirPods \cite{saferemr}.

\begin{figure}
\centering
\begin{subfigure}[b]{\linewidth}
\centering
\includegraphics[width = \linewidth]{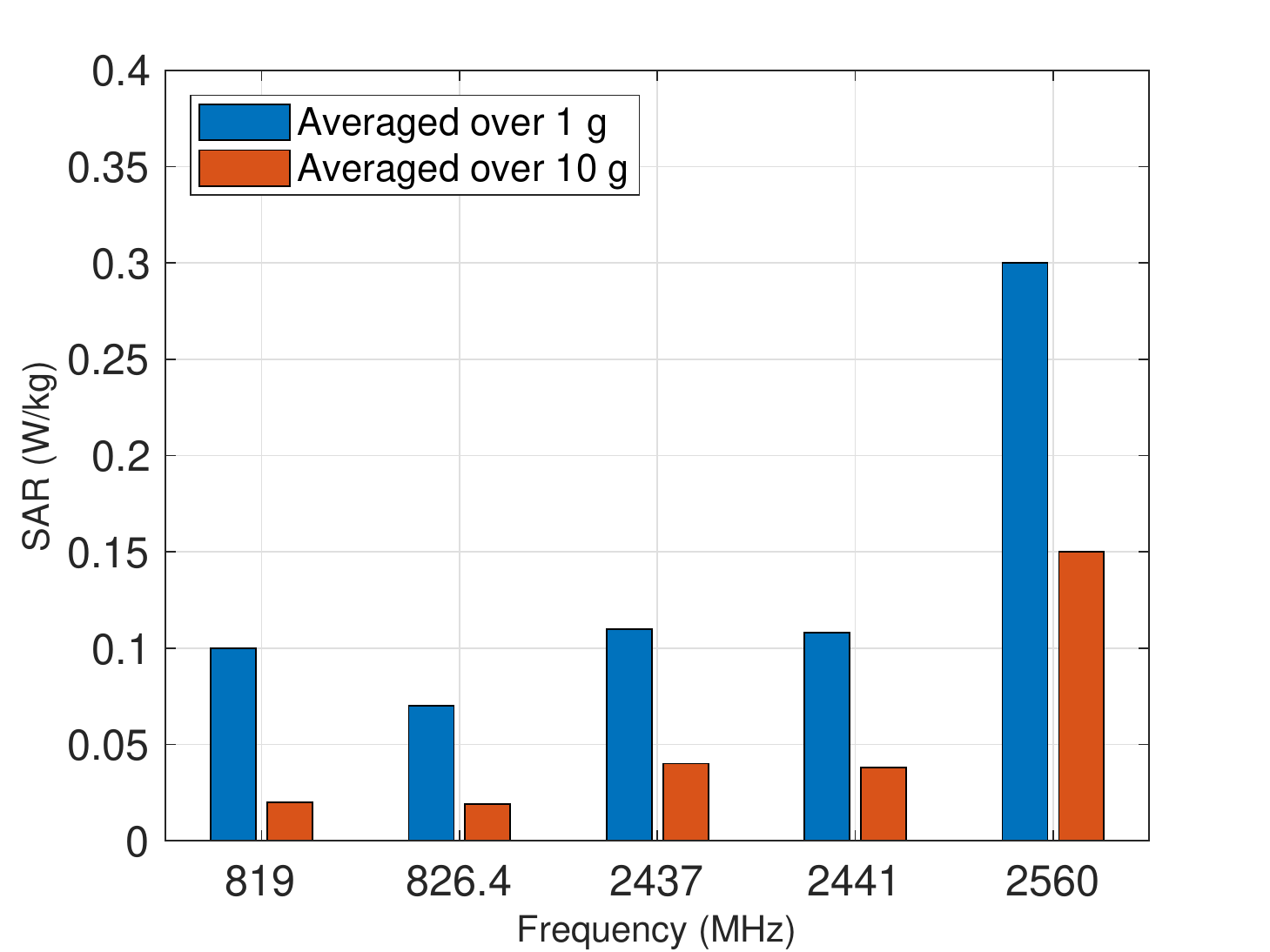}
\caption{Apple A1889}
\label{fig_a1889}
\end{subfigure}\hfill
\begin{subfigure}[b]{\linewidth}
\centering
\includegraphics[width = \linewidth]{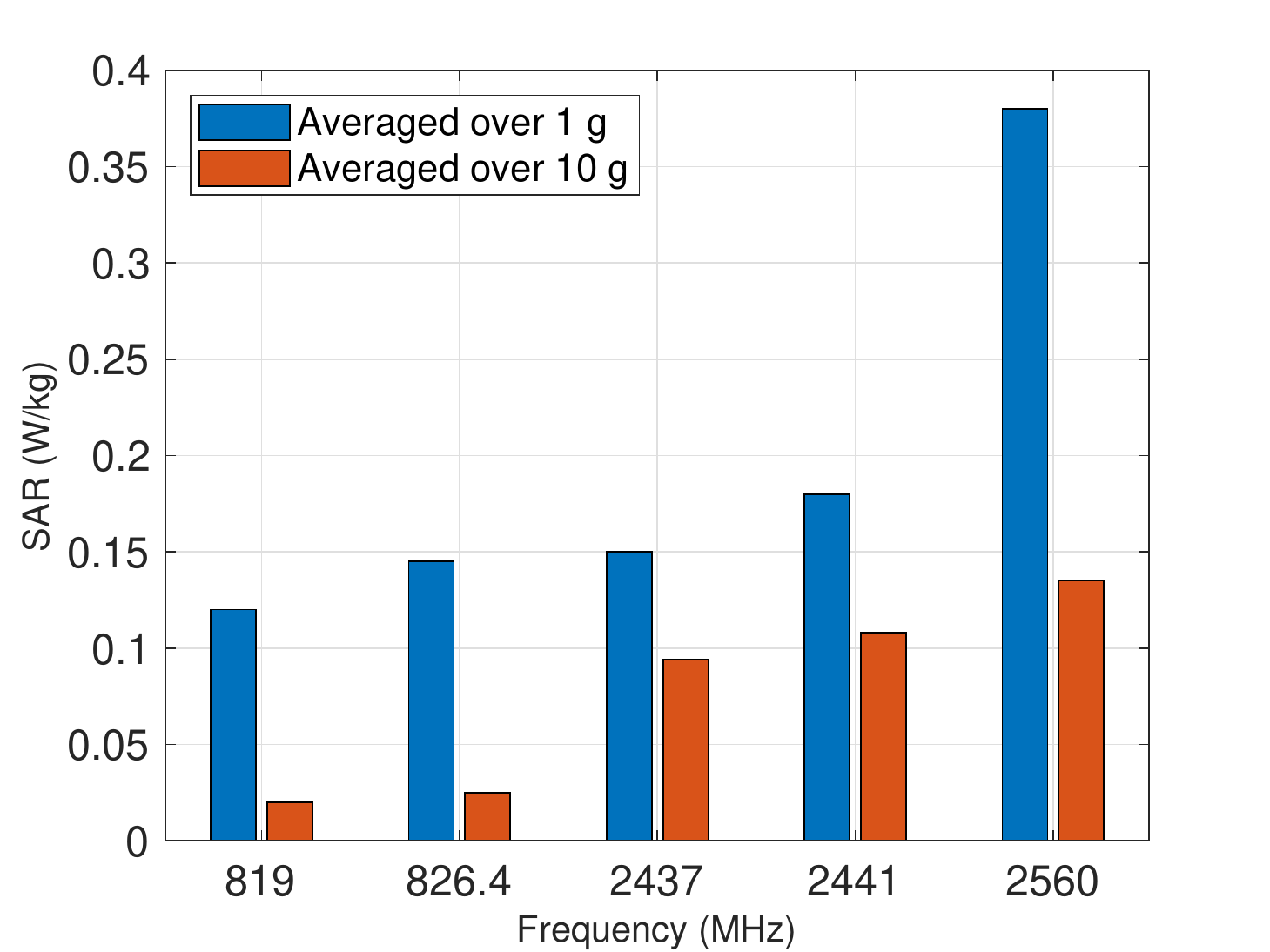}
\caption{Apple A1891}
\label{fig_a1891}
\end{subfigure}
\caption{SAR versus frequency for Apple A1889 and A1891}
\label{fig_apple}
\end{figure}

\subsection{Apple Watch}
Table \ref{table_list2} SAR and the other parameters for Apple Watch series, which operate in a wider selection of spectrum bands--\textit{i.e.}, UMTS at 850 and 1750 MHz, LTE in Band 7 and 26, IEEE 802.11b, and Bluetooth.

During testing, Apple Watch radios were set to their highest transmission levels and placed in positions that simulate a normal use with 10 mm separation, for a use against the head, and no separation, for one against the wrist \cite{appleexposure}.

Looking across the Apple Watch Series 4 models, the highest SARs are found as 0.37 W/kg, 0.17 W/kg, and 0.13 W/kg, in cellular transmission, Wi-Fi, and Bluetooth, respectively. The model also supports simultaneous transmission (\textit{i.e.}, cellular plus Wi-Fi), which yields the SAR of 0.50 W/kg. Notice that all of these four values fulfill the current safety guidelines. (Recall 1.6 W/Kg and 2.0 W/Kg by the FCC \cite{fcc01} and ICNIRP \cite{ICNIRP74}, respectively.)

As an example, Figs. \ref{fig_a1889} and \ref{fig_a1891} illustrate the SAR of Apple A1889 and A1891 versus the carrier frequency. Commonly from the two figures, one can observe a general pattern showing that the SAR increases as the frequency gets higher. The rationale can be recalled from the definition of a SAR, which has the `penetration depth' in its denominator: a higher carrier frequency yields a smaller penetration depth. In other words, although the EMF with a higher frequency travels to a shallower point into human skin, a greater amount of EMF energy is absorbed.

\section{Human EMF Exposure in Werable Communications Operating in mmW Spectrum}\label{sec_mmw}
\subsection{Motivation}
The potential of mmW frequencies for wearable communications is enormous for applications requiring Gbps throughput. Such networks might use wireless standards including 5G \cite{jsac} or IEEE 802.11ad/ay \cite{hst}, based on which commercial products are already available.

As have been observed in Eq. (\ref{eq_sar_dphi}), the major concern regarding wearable communications in mmW is higher SAR due to the extremely small penetration depth, $\delta$, as shown in Eq. (\ref{eq_sar_dphi}).

Moreover, some tissues (\textit{e.g.}, eyes) are especially vulnerable to mmW radiation-induced heating and require more attention \cite{vtmag18}. It is necessary to continually update regulations based on new materials, frequencies, device types, and transmitted powers. Additionally, manufacturers must be educated with the newest research/regulations to better address consumer concerns and promote this new technology.

Out of various spectrum bands, this paper focuses unlicensed bands, which enable cheaper and less complex devices (as well as longer battery life), all of which are desirable for wearable communications and networking \cite{vtmag18}. The 2.4 GHz industrial, scientific, and medical (ISM) band has already long been coveted by numerous unlicensed communications systems. The FCC recently opened up an additional 7 GHz of spectrum available for unlicensed use through 64-71 GHz, which now provides a historic 14 GHz of contiguous chunk through 57-71 GHz--also known as the 60 GHz band \cite{jsac}.

Therefore, this paper considers a wearable communications environment with two different carrier frequencies--\textit{i.e.}, 2.4 GHz and 60 GHz. It enables a comparison of the impacts of carrier frequency on the human EMF exposure. Adopting two different operating frequencies differentiates wearable communications environments. The adopted parameters for the two scenarios are summarized in Table \ref{table_parameters}.

\subsection{Antenna Pattern}
We adopt the general antenna pattern equation given by \cite{jsac}
\begin{align}\label{eq_pattern_wifi}
G\left(\theta\right)=G_{max} - \exp\left(-2\pi j\delta \sin\theta\right) \rm{~[dB]}
\end{align}
where $\delta$ denotes the antenna element separation distance, and $\theta$ denotes a general angle. Notice that both azimuth and elevation angles are assumed to affect the antenna gain based on Eq. (\ref{eq_pattern_wifi}). The parameters $\theta_{3db}$ and $A_m$ for a wearable environment are considered as 93$^\circ$ \cite{cosan} and 30 dB, respectively.

It is more desirable to assume a continuous line-of-sight (LOS) \cite{heath_presentation_2015} link between a wearable device and the human body. Based on this rationale, this paper adopts a free space path loss (FSPL) model for calculation of a path loss, which is formally written as
\begin{align}\label{eq_fspl}
\text{PL} = 20\log(d) + 20\log(f) - 27.55 \rm{~[dB]}
\end{align}
where \textit{d} (m) is the distance between the antennas and \textit{f} (MHz) represents the operating frequency.

\begin{table}[t]
\small
\caption{Parameters for on-body wearable communications}
\centering
\begin{tabular}{|c|c|c|c|c}
\hline 
\textbf{Parameter}{\cellcolor{gray!20}} & 60 GHz {\cellcolor{gray!20}} & 2.4 GHz {\cellcolor{gray!20}}\\ \hline
\hline
Bandwidth & 2.16 GHz \cite{heath_presentation_2015} & 93 MHz \cite{cosan}\\
Max antenna gain & 11.9 dBi \cite{chahat2012} & 10.1 dBi\\
Transmit power & 10 dBm & 2 dBm \cite{wagih_2019}\\
Antenna elements & 16 \cite{heath_presentation_2015} & 4\\
\hline
Path loss model & \multicolumn{2}{c|}{Free space path loss (FSPL)}\\
Receiver gain & \multicolumn{2}{c|}{10 dBi}\\
Receiver noise figure & \multicolumn{2}{c|}{6 dB}\\
Temperature & \multicolumn{2}{c|}{290 K}\\ \hline
\end{tabular}
\label{table_parameters}
\end{table}

\begin{figure}
\centering
\begin{subfigure}[b]{\linewidth}
\centering
\includegraphics[width =\linewidth]{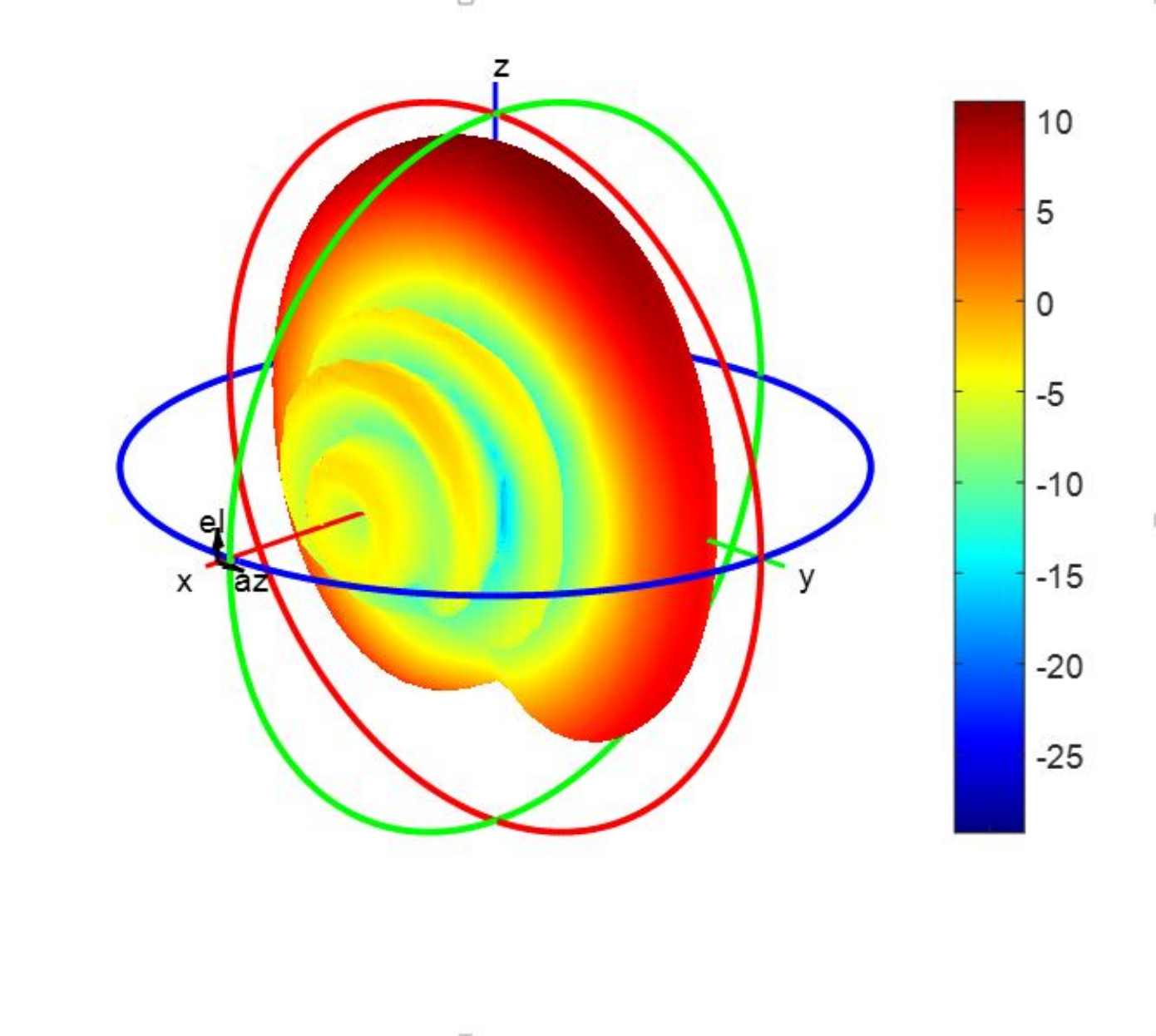}
\caption{At 2.4 GHz}
\label{2_4pattern}
\end{subfigure}\hfill
\begin{subfigure}[b]{\linewidth}
\centering
\includegraphics[width =\linewidth]{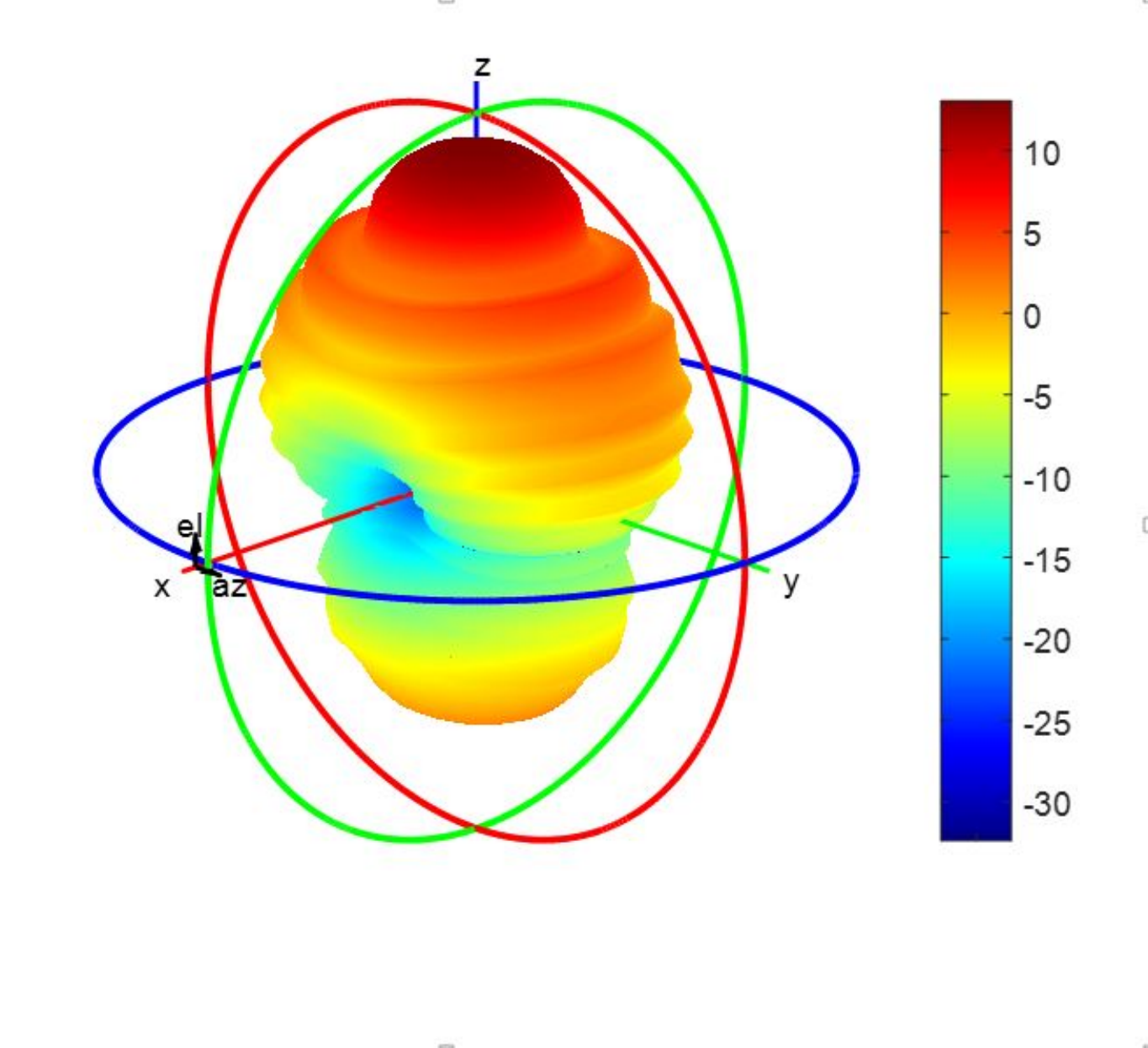}
\caption{At 60 GHz}
\label{60_pattern}
\end{subfigure}
\caption{Example antenna patterns for wearable communications}
\label{fig_pattern}
\end{figure}

\subsubsection{Wearable Antenna at 2.4 GHz}
Different communication links require different radiation patterns for the wearable operation. When two on-body devices communicate, an omni-directional pattern seems to meet the requirements. However, when an on-body sensor communicates with an off-body device, a broadside pattern (or patch-like pattern) is more desirable. To support these two operating states, a patch like reconfigurable pattern has been reported \cite{yan2016} where the antenna can change its resonance to support both states with the same frequency.

The problem with EMF transmissions at 2.4 GHz is more ``spread,'' not just towards the intended recipient. As such, for 2.4 GHz, this paper assumes a microstrip patch antenna which is more of omni-directional in pattern on a certain plane as demonstrated in Fig. \ref{2_4pattern}.

\subsubsection{Wearable Antenna at 60 GHz}
It is the opposite in communications at 60 GHz. To overcome high attenuation, a communications system in the band usually adopts a ``highly directive'' radiation pattern.

This matches the need in wearables. Commonly, wearable antennas are expected to be low-profile, lightweight, compact, and conformable to the body shape. As such, at 60 GHz, an end-fire pattern has been proposed \cite{chahat2012}. In such a pattern, the direction of maximum radiation can be set parallel to the body surface. Also, a microstrip-fed Yagi-Uda antenna meets these specifications \cite{chahat2012}. Fig. \ref{60_pattern} shows the antenna pattern of a Yagi-Uda antenna operating at 60 GHz which shows more of a directive radiation pattern compared to the one used at 2.4 GHz. We assumed 16 directors for the 60 GHz pattern as described in Table \ref{table_parameters}.

\begin{figure}
\centering
\begin{subfigure}[b]{\linewidth}
\centering
\includegraphics[width = \linewidth]{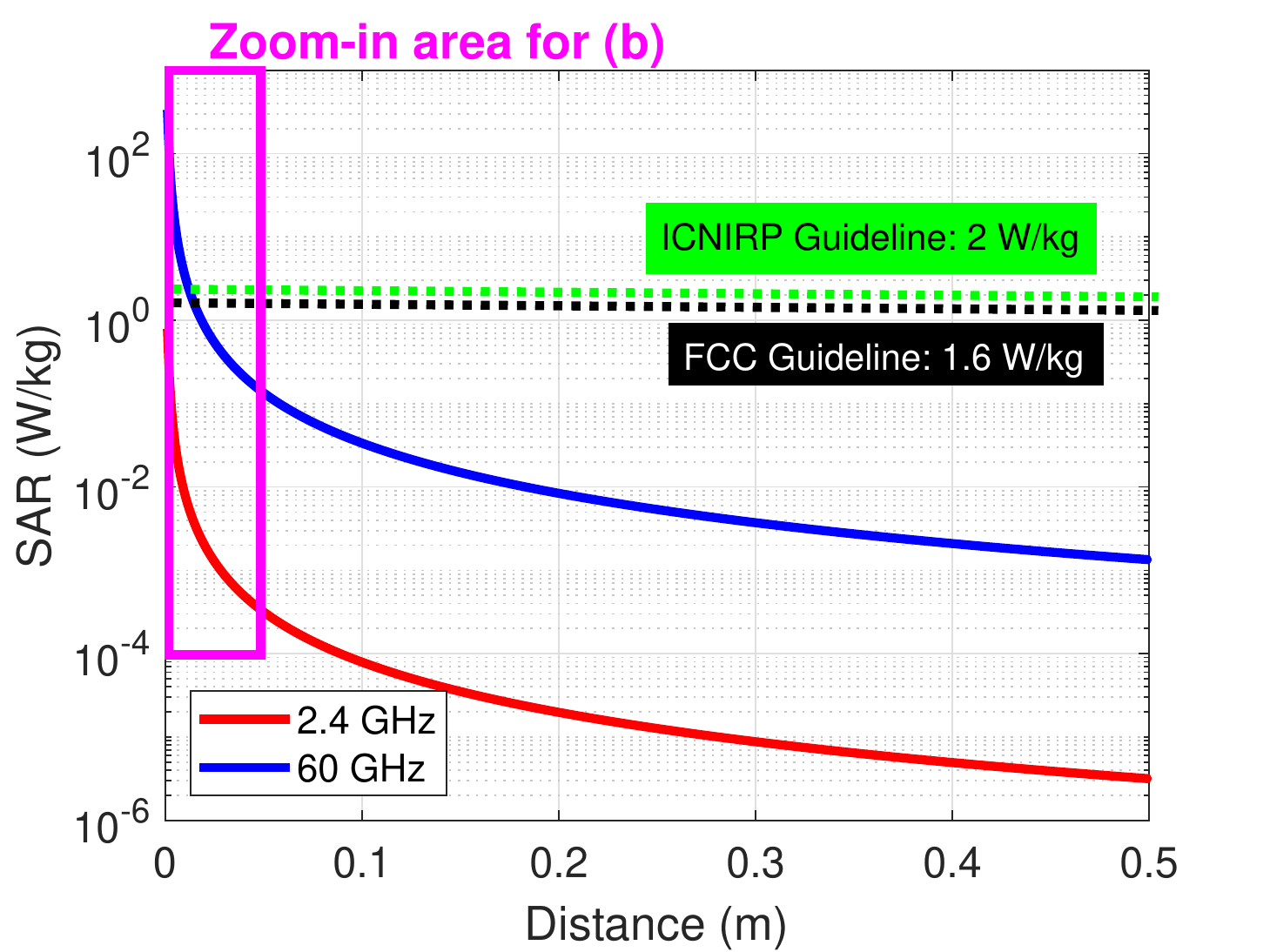}
\caption{SAR versus distance}
\label{fig_sar_original}
\end{subfigure}\hfill
\begin{subfigure}[b]{\linewidth}
\centering
\includegraphics[width = \linewidth]{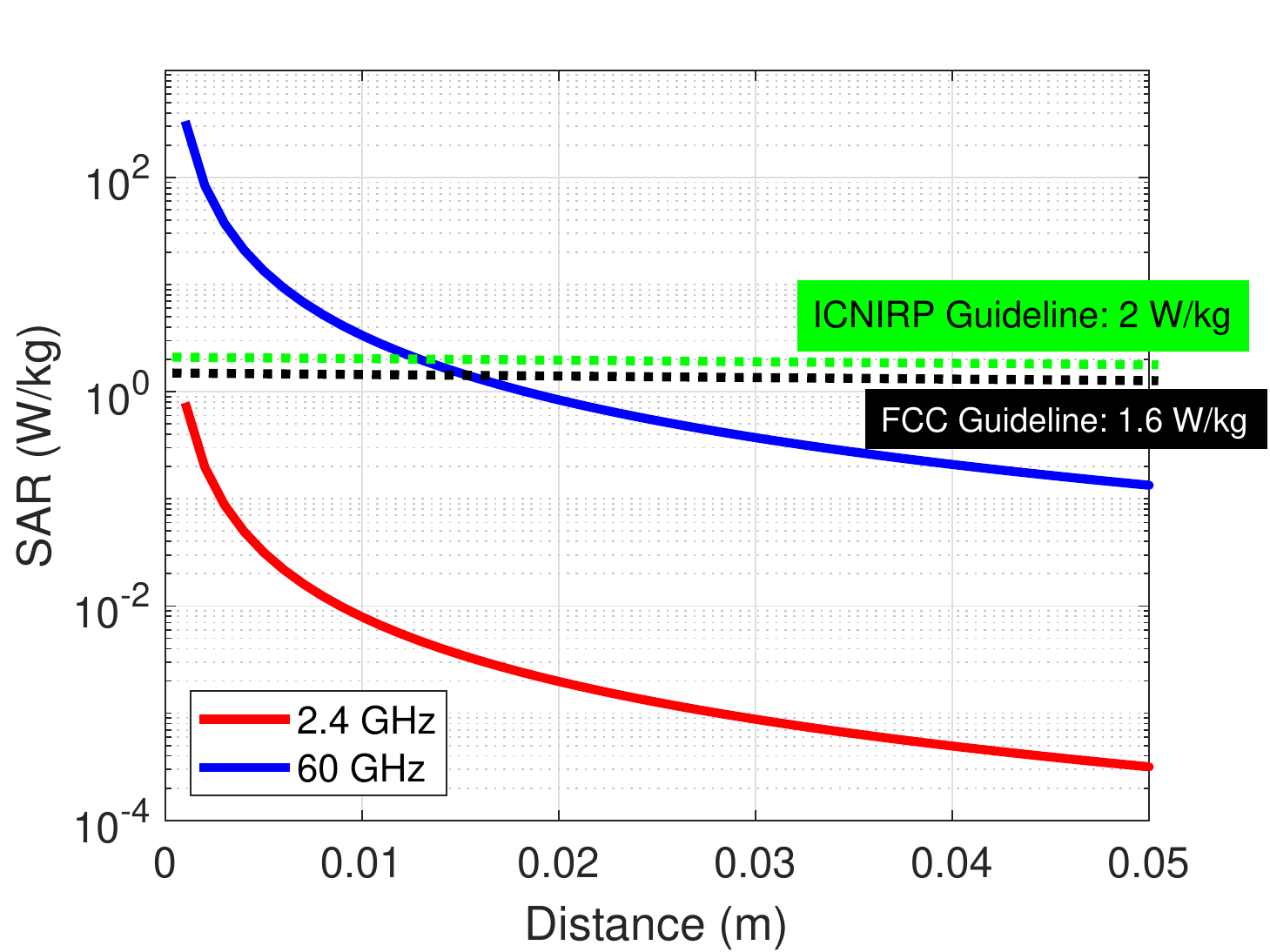}
\caption{A zoom-in view of (a) within [0,5] cm}
\label{fig_sar_zoom}
\end{subfigure}
\caption{SAR comparison between 2.4 GHz and 60 GHz}
\label{fig_sar}
\vspace{0.2 in}
\end{figure}

\subsection{SAR Evaluation}
Fig. \ref{fig_sar} represents the human EMF exposure from wearable devices in terms of SAR. The result suggests that the average SAR measured at a certain distance from a wearable device can exceed the existing ICNIRP or FCC guidelines.

Fig. \ref{fig_sar_zoom} shows a zoomed-in display for the range of [0, 5] cm of Fig. \ref{fig_sar_original}. It provides a closer investigation of the minimum safe distance between the wearable devices that are mounted on the garments of the human body and the human skin. It can be suggested from Fig. \ref{fig_sar_zoom} that for a wearable device operating at 60 GHz, a separation distance of 12 and 15 mm are required according to the current ICNIRP and FCC guidelines, respectively. However, the SAR for operating frequency at 2.4 GHz remains far below the guideline from the very first point.

Although one can consider this distance to be very minimal, for a wearable environment, this small distance should also be taken into account for designing wearable devices on the human body. For instance, if a soldier is wearing an on-body device that is mounted on his smart glasses or the VR helmet, its radiation can impact sensitive organs like the eyes or the human brain. The adaptation of (i) higher number of antenna directors and (ii) higher transmit power are the primary reasons for the elevation of this human EMF exposure at 60 GHz. As noted previously, this elevation in SAR can increase the temperature at the surface of the human skin which may have a lethal impact on the human body when dosed in a continued manner or over a long-term period \cite{temperature}.

\section{Conclusions}
This paper has presented (i) a mathematical framework for SAR analysis, (ii) state-of-the-art standards and measurement methodologies, (iii) exact SAR levels of currently sold wearable devices, and (iv) simulation results of SAR levels evaluated at 60 GHz. The mathematical framework informed how exactly the SAR is measured: \textit{i.e.}, definition of SAR, derivation of SAR according to different variables, and an average over time. Then, this paper showed how the time-averaged SAR is actually measured in the current SAR measurement methodologies for standard compliance tests. Based on these mathematical expressions and measurement methodologies, some popular commercial products were investigated on their SAR levels. Considering the latest paradigm shift to mmW communications, this paper evaluated the SAR measured at 60 GHz and compared it to that at 2.4 GHz. Although the current guidelines do not regulate SAR at 60 GHz, inferring the guidelines defined at lower frequencies, wearable communications at 60 GHz were found to cause SAR exceeding the guidelines. The separation distances were 12 and 15 mm, according to ICNIRP and FCC, respectively.

The suggestions of this paper are clear: in order to better mitigate consumer concerns and promote this new technology, we urge to keep (i) the general public informed of the latest information on the commercial wearable devices; (ii) the related safety regulations up to date; and (iii) the manufacturers educated about the newest research and regulations.

\end{document}